\documentclass[showpacs,superscriptaddress,floatfix,a4paper,twocolumn,prb,aps]{revtex4-1}
\usepackage[pdftex]{graphicx}
\usepackage{amsfonts}
\usepackage{amsmath}
\usepackage{amssymb}
\usepackage{upgreek}
\usepackage{dsfont}
\usepackage{bm}
\usepackage{color}
\usepackage{hyperref}
\usepackage{times}

\begin{document}
\title{Probing surface states exposed by crystal terminations at arbitrary orientations of three-dimensional topological insulators }
\author{Sthitadhi Roy}
\affiliation{Max-Planck-Institut f\"ur Physik komplexer Systeme, N\"othnitzer 
Stra$\beta$e 38, 01187 Dresden, Germany}
\author{Kush Saha}
\affiliation{D\'epartement de Physique and Regroupement Qu\'eb\'ecois sur les Mat\'eriaux de Pointe, Universit\'e de Sherbrooke, Sherbrooke, Qu\'ebec, Canada J1K 2R1}
\author{Sourin Das}
\affiliation{Department of Physics and Astrophysics,
University of Delhi, Delhi 110 007, India}
\affiliation{Max-Planck-Institut f\"ur Physik komplexer Systeme, N\"othnitzer 
Stra$\beta$e 38, 01187 Dresden, Germany}

\begin{abstract}
The topological properties of the bulk band structure of a three-dimensional topological insulator (TI) manifest themselves in the form of metallic surface states. In this paper, we propose a probe which directly couples to an exotic property of these surface states, namely the spin-momentum locking. We show that the information regarding the spin textures, so extracted, for different surfaces can be put together to reconstruct the parameters characterizing the bulk band structure of the material, hence acting as a hologram. For specific TI materials like, $\text{Bi}_2\text{Se}_3, \text{ Bi}_2\text{Te}_3 \text{ and Sb}_2\text{Te}_3$, the planar surface states are distinct from one another with regard to their spectrum and the associated spin texture  for each angle ($\theta$), which the normal to the surface makes with the crystal growth axis. We develop a tunnel Hamiltonian between such arbitrary surfaces and a spin polarized STM which provides a unique fingerprint of the dispersion and the associated spin texture corresponding to each $\theta$.  Additionally, the theory presented in this article can be used to extract value of $\theta$ for a given arbitrary planar surface from the STM spectra itself hence effectively mimicking X-ray spectroscopy. 
\end{abstract}

\maketitle

\section{Introduction}\label{sec:intro}
Topological insulators (TI) have been a very active subject of research in condensed matter physics since their inception.\cite{Kane2005,Kane2005a,Bernevig2006,Bernevig2006a,Konig2007,Fu2007,Moore2010,Hasan2010,Qi2011,Bernevig2013book} The discovery of strong 3D TI materials\cite{Fu2007,Zhang2009,Chen2009,Moore2010,Hasan2010,Qi2011,Bernevig2013book} which have an insulating bulk and topologically protected metallic surface states has led to an ever increasing amount of studies both theoretically and experimentally. These materials are known to exhibit exotic properties which are attributed to the topological aspect of the bands associated with these materials.\cite{Fu2007,Moore2007,Roy2009,Zhang2009} TIs are what can be called ÒholographicÓ materials, in the sense that the properties of the topological gapped bulk can be deduced from the image of these topological properties on the surface states. A special feature of surface states in these materials is the presence of spin-momentum locking. The central motivation of this paper is to propose a holographic probe for 3D TIs which gives a read out of the parameters characterizing the bulk by studying the spin textures of many different surfaces of the TI, corresponding to many different cleaving angles with respect to the crystal growth axis, within an electrical transport setup in a non-invasive fashion.

The spin textures of the surface states of TIs have been studied by spin-resolved angle resolved photoemission spectroscopy (ARPES)\cite{Hsieh2008,Chen2009,Xia2009,Hasan2014} or by scanning tunneling microscope (STM)\cite{Roushan2009,Zhang2009a,Alpichshev2010,Saha2011,Khanna2013} via  study of quasi-particle interference patterns induced by weak disorder potential. Though spin-resolved ARPES serves as a very efficient probe for scanning the surface state spectrum of 3D TI leading to a complete reconstruction of the spin texture of the Fermi surface, it does not provide information regarding influence of spin-momentum locking in electrical transport properties of the surface state. On the other hand, STM can be used to gather information regarding spin-momentum locking via transport but only in the disordered limit as it relies on quasi-particle interference\cite{Roushan2009} which is produced due to disorder-induced scattering and the information hence extracted is indirect. Also, experiments with multiple ferromagnetic contacts pad\cite{Yokoyama2010,Taguchi2014,Li2014a,Liu2014,Dankert2014} do provide a strong indications of spin-momentum locking but at the same time they are far from being comprehensive as far possibility of reconstruction of the spin texture of the Fermi surface is concerned. Hence an electrical transport probe which works in the ballistic regime and probes the spin texture as directly as the spin-resolved ARPES with minimal invasion into the TI surface state is desirable.\cite{Roy2014}

Further note that  the technique of spin-polarized ARPES works very well as long as the spin-momentum locking is between a physical electron spin and its momentum, on the other hand if the locking is with a pseudo spin degree of freedom like the case of $\theta \neq 0, \pi$\cite{Zhang2012,Zhang2013} then the expected success of spin-polarized ARPES in determining spin texture is a {\it priori} not clear \cite{Silvestrov2012}. In this article we carry out an extensive study on the possibility of using spin polarized STM as a probe for the surface state corresponding to arbitrary value of $\theta$ in the ballistic limit. We exploit the new concept of a multi-terminal tunnel magnetoresistance (TMR)\cite{Slonczewski1989} response defined in Ref.~\onlinecite{Roy2014} to develop a strategy for probing the Fermi surface and their spin texture for surface states corresponding to arbitrary values of $\theta$. For $\theta\neq 0,\pi$, the degree of freedom which couples to the momentum is a surface dependent linear combination of two SU(2) degrees of freedom, the electron spin and the orbital pseudo spin, however the spin polarized STM has only the electron spin as its degree of freedom. So, in the process we come up with a tunnel Hamiltonian for such junctions, which is a crucial development in itself. We theoretically show that the proposed strategy indeed provides enough information to facilitate reconstruction of the Fermi surface and the spin textures. The central outcome of our study lies in the fact that, when the collection of information extracted from the STM scan of surfaces correspond to all possible $\theta$  is put together, it facilitates identification of fundamental parameters of the bulk band structure and hence effectively acting  like a hologram.

The rest of the paper is organized as follows: in Sec.~\ref{sec:schematic}, the geometry and the setup is discussed. In Sec.~\ref{sec:model}, the results of Ref.[\onlinecite{Zhang2012}] regarding the surface states and their spin texture is briefly reviewed followed by discussions on degeneracy induced breaking of symmetries  by surface states in Sec.~\ref{sec:degeneracy}.  In Sec.~\ref{sec:stm}, a model for a spin polarized  STM  is presented and the corresponding momentum resolved current is discussed. Current from STM injected into the TI surface is calculated to leading orders perturbatively in the weak tunnel coupling limit. In Sec.\ref{sec:spintexture}, we discuss the strategy for reconstructing the spin texture corresponding to different surfaces via a detailed study of tunneling current which carry unique fingerprints of the corresponding Fermi surfaces and its spin textures. In Sec.\ref{sec:hologram} we show that the spin textures for different surfaces can be used to extract fundamental parameters of the bulk Hamiltonian which effectively determine the surface dependent Fermi velocities for different $\theta$, hence acting as a hologram of the bulk. Finally the results are summarized in Sec.~\ref{sec:conclusion} along with outlook.
\section{Experimental geometry}\label{sec:schematic}
\begin{figure}[t]
\begin{center}
\includegraphics[width=0.69\columnwidth]{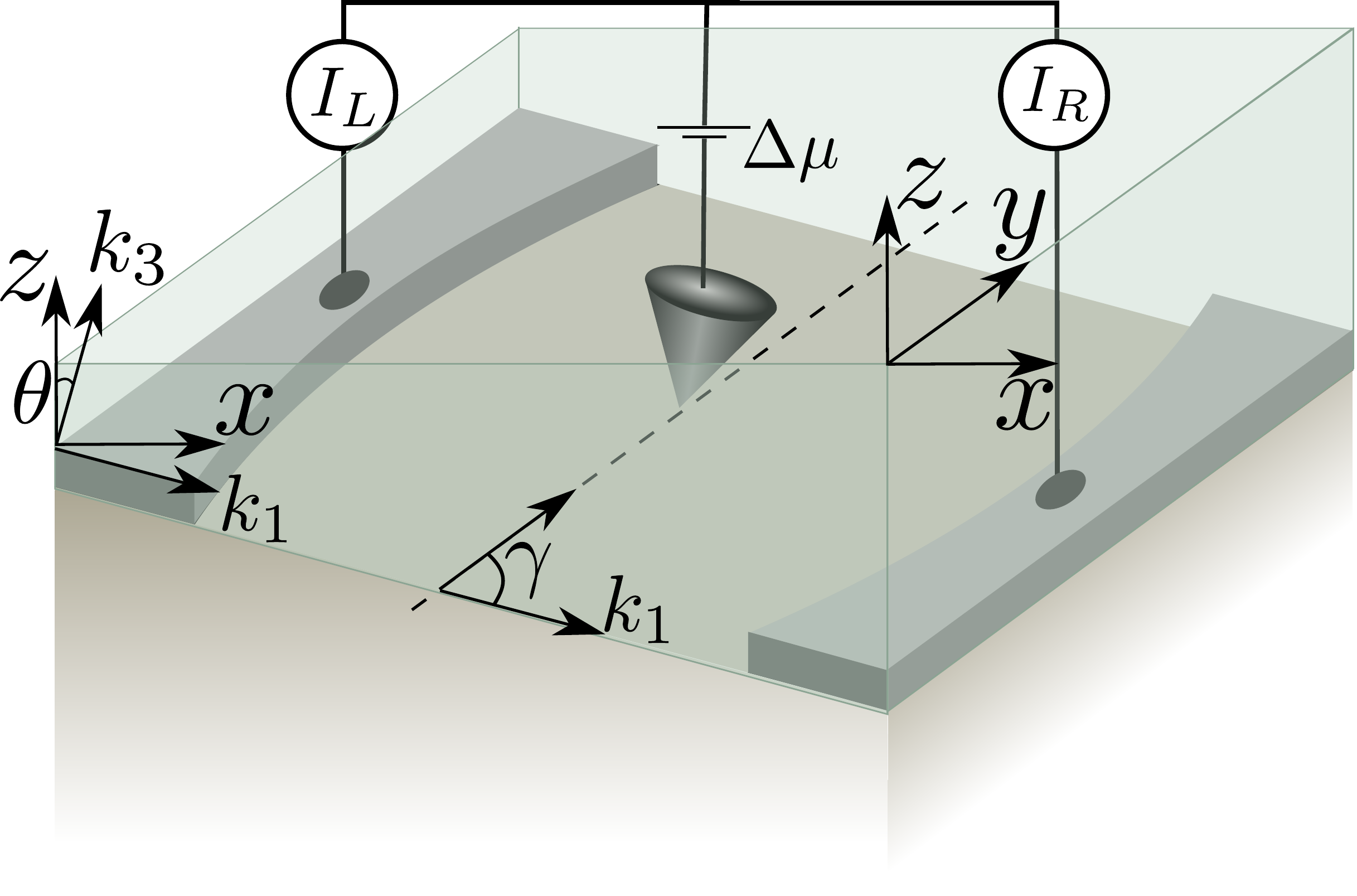}
\includegraphics[width=0.29\columnwidth]{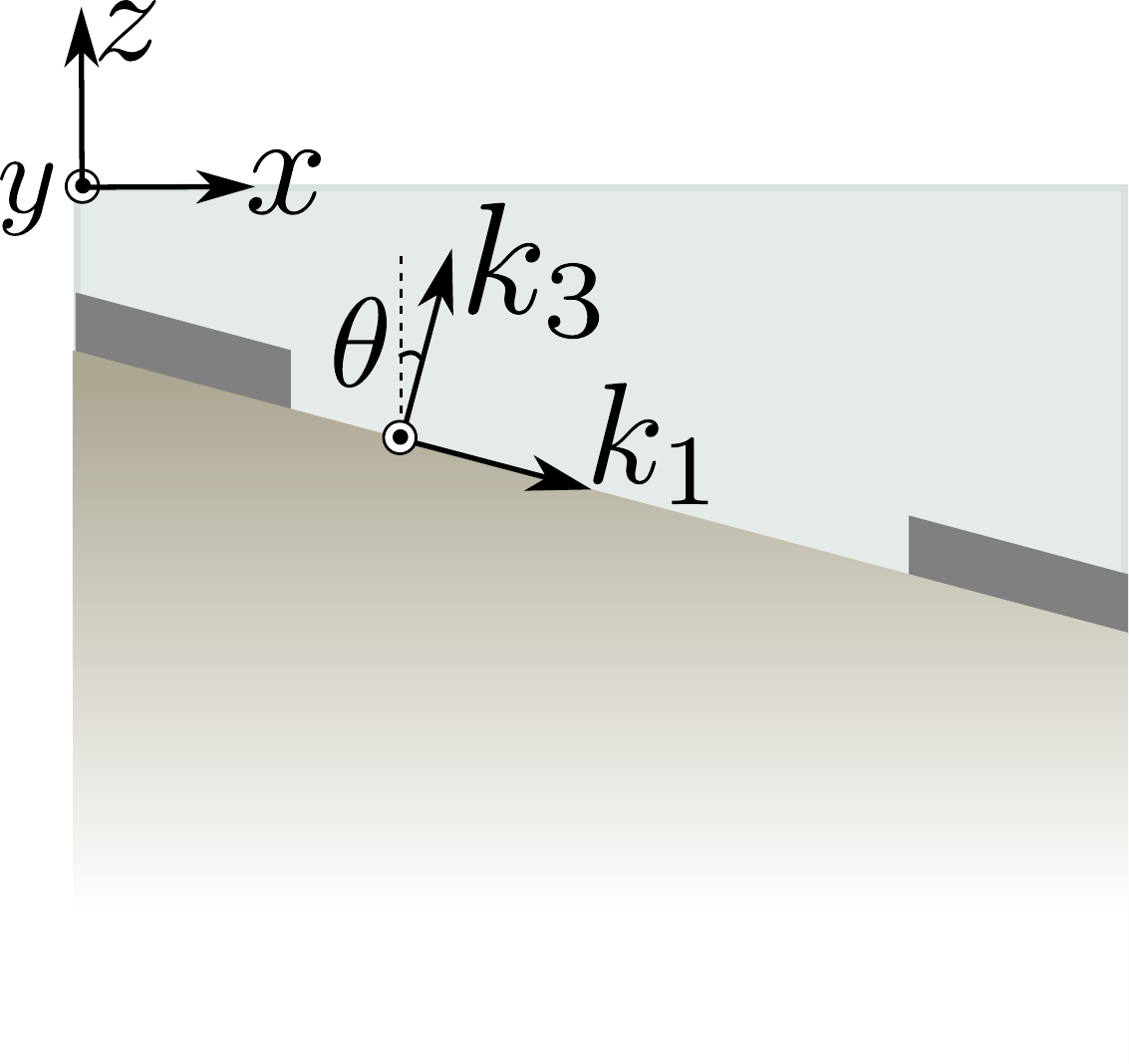}
\end{center}
\caption{Left:A schematic of the set-up involving an arbitrarily cut surface of the TI, STM and the two contact pads. The surface shown with the contacts has its normal (pointing along $\hat{k}_3$) at an angle $\theta$ to the crystal growth axis (along $\hat{z}$). As a reference, a translucent surface lying in the $x$-$y$ plane is shown which is perpendicular to the $z$-axis. For more details refer to the text in Sec.\ref{sec:schematic}. Right: The schematic setup viewed along the $y$-direction clearly shows the surface of interest along with the coordinate system. }
\label{fig:schematic}
\end{figure}

In this section, the schematic of the setup is discussed along with setting up the conventions and notations which are used in the rest of the paper. A schematic diagram of the setup is shown in Fig.~(\ref{fig:schematic}). The crystal growth axis is always taken along the $z$ direction, and an arbitrary surface denoted by $\Sigma(\theta)$ is exposed by taking a cut parallel to the $y$ direction, such that the normal to the surface makes an angle $\theta$ with the crystal growth axis. Momentum normal to the exposed surface is denoted by $k_3$. Since the cut is parallel to the $y$ axis, hence one of the in-plane momentum remains $k_y$ as we vary $\theta$. The other in-plane momentum orthogonal to $k_y$ is denoted by $k_1$ which is basically $k_x$ rotated by an angle $\theta$ about the $y$ axis as shown in Fig.~(\ref{fig:schematic}). For convenience of the readers we have kept the coordinate system defining the surface same as in Ref.~[\onlinecite{Zhang2012}]. Two contact pads are placed on diagrammatically opposite sides of the exposed surface such that each contact collects the current injected in that half of the sample in which the contact pad is placed. An imaginary line which divides the sample into two halves is shown by the dashed line in Fig.~(\ref{fig:schematic}). The angle it makes with the $k_1$ axis is denoted by $\gamma$. According to this convention the current collected in the left contact denoted by $I_L$ is the current that flows in the region covered by the range of angle  from $\gamma$ to $\gamma+\pi$, whereas the current collected by the right contact $I_R$, is the one collected in the region spanned by angle ranging from $\gamma+\pi$ to $\gamma +2\pi$. In this way, a strict sense of consistency is maintained in regard to what is $I_L$ and $I_R$. The current asymmetry $\Delta I$ is defined as the difference of the two, i.e,  $\Delta I = I_L-I_R$ and the total current injected from the tip into the TI surface  is denoted by $I_0=I_L+I_R$.
In our analysis, we neglect the edge states that could in principle, appear at the edges between two surfaces with different orientation of a 3D TI.\cite{Deb2014,Zhang2013}

\section{Surface states and spin textures}\label{sec:model}
Band structure studies of 3D TI materials like $\text{Bi}_2\text{Se}_3, \text{ Bi}_2\text{Te}_3 \text{ and Sb}_2\text{Te}_3$ are such that the low energy sector of the theory expanded around the center of the Brillouin zone ($\Gamma$-point) are spanned by states labeled by only two quantum numbers.\cite{Zhang2012} One being the electron spin ($\sigma$) quantum number and the second one is a parity quantum number ($\tau$) owing to inversion symmetry of the unit cell and it takes $\pm 1$ as its eigenvalue. Hence, with spin and parity put together, the effective low energy theory for these materials reduces to a four band model. From here on we will refer to the space spanned by eigenstates of $\tau$ as the orbital pseudo-spin space. For the surface perpendicular to the crystal growth axis ($\theta=0 \text{ and } \pi$), the two sectors belonging to the spin and the parity remain completely decoupled as far as dispersing degree of freedom are concerned  but this fact is true only for  $\theta=0 \text{ and } \pi$ surface. In an elegant work by Zhang {\it et.al}.,\cite{Zhang2012} the physics of surfaces oblique to the crystal growth axis was considered. It was shown that the momentum modes of these surface states coupled to new $SU(2)$ degree of freedom which is different from the spin. These degree of freedoms are constructed out of combinations of spin and the parity. The exact composition of spin and parity contribution depends directly on the angle $\theta$. Furthermore, unlike the top surface, the Fermi surface is no more circular but elliptic where the eccentricity again depends on the angle $\theta$.  

We start with a review of the expressions concerning the surface states for arbitrary $\theta$ obtained in Ref.~[\onlinecite{Zhang2012}] which are important for evaluating the tunneling current from a spin polarized STM into the TI surface.  A self-contained reproduction of the expressions of Ref.~[\onlinecite{Zhang2012}] can be found in Appendix \ref{sec:app}.

The low energy Hamiltonian expanded around the $\Gamma$-point  for 3D TI material like $\text{Bi}_2\text{Se}_3$ retained up to linear order in momentum is given by
\begin{equation}
 \mathcal{H}= (-m_0\tau_z + v_zk_z\tau_y) \otimes \mathds{I}_{\sigma}  + v_{\|}~  \tau_x \otimes (k_y\sigma_x - k_x\sigma_y),
 \label{eq:bulkham}
\end{equation}
where the basis chosen is given by $(\vert+\uparrow\rangle,\vert+\downarrow\rangle,\vert-\uparrow\rangle,\vert-\downarrow\rangle)$. Here $\pm$ denotes the even and odd parity orbitals and the $\uparrow/\downarrow$ denotes the $z$-component of electron spin where $z$-axis is taken to be parallel to the crystal growth axis and $m_0$ denotes the bulk band gap. This basis renders the surface Hamiltonian block diagonal in the parity space and helps in motivating the formalism for calculating the tunneling current from a spin-polarized STM into two channels corresponding to the two parity orbitals. Following Ref.~[\onlinecite{Zhang2009}], we use the parameter values $v_z=2.2 \text{ eV\AA}$ and $v_\|=4.1 \text{ eV\AA}$. To derive the surface Hamiltonian for an arbitrary surface $\Sigma(\theta)$ that is exposed by cleaving the crystal at an angle $\theta$ with the crystal growth axis, it is convenient to work with a  local surface dependent frame of reference spanned by $\hat{k_1},\hat{k_y},\hat{k_3}$ axes (see Fig.\ref{fig:schematic}), where $\hat{k_1},\hat{k_y}$ are the {\it in-plane vectors}. The rotation of the plane is always done about the $\hat{k_y}$ axis and $\hat{k_3}$ is chosen to be perpendicular to the given surface. In such a frame, the Hamiltonian in Eq.~(\ref{eq:bulkham}) can be written as
\begin{equation}
 \mathcal{H} = -m_0 T_z + (v_3k_3+v_0k_1)T_y + (v_{\|}k_yS_x - v_1k_1S_y)T_x,
 \label{eq:bulkhamlocal}
\end{equation}
where $\bm{S}$ and $\bm{T}$ are pseudospins defined by surface dependent linear combinations of products of  $\bm{\tau}\otimes\mathds{I_{\sigma}}$ and $\mathds{I_{\tau}}\otimes\bm{\sigma}$. They are given by\cite{Zhang2012} 
\begin{equation}
\begin{split}
 \bm{T}&=\{\alpha\tau_x\otimes \mathds{I}_{\sigma} +\beta\tau_y\otimes \sigma_y,\alpha\tau_y\otimes \mathds{I}_{\sigma} -\beta\tau_x\otimes \sigma_y,\tau_z \otimes \mathds{I}_{\sigma}\}\\
 \bm{S} &= \{\alpha  \mathds{I}_{\tau}\otimes \sigma_x-\beta\tau_z\otimes \sigma_z, \mathds{I}_{\tau}\otimes \sigma_y, \alpha \mathds{I}_{\tau}\otimes \sigma_z+\beta\tau_z\otimes\sigma_x\},
 \end{split}
\end{equation}
where $v_3=\sqrt{(v_{\|} \sin\theta)^2+(v_{z}\cos\theta)^2}$, $\alpha=v_z \cos\theta/v_3$, and  $\beta=v_{\|} \sin\theta/v_3$. Also $v_0 = (v_{\|}^2-v_z^2)\sin\theta\cos\theta/v_3$ and $v_1=v_zv_{\|}/v_3$. Both $\bm{T}$ and $ \bm{S} $ independently satisfy the $SU(2)$ algebra and they also satisfy $[\bm{T}_{i},  \bm{S}_{j}]=0$ for all $i,j$. Using the topological boundary condition\cite{Jackiw1976} given in Ref. (\onlinecite{Zhang2012}) the surface Hamiltonian is obtained as    
\begin{equation}
 \mathcal{H}_{\text{surf}}(\theta)=v_{\|}k_yS_x - v_1k_1S_y.
 \label{eq:surfham}
\end{equation}
In the basis chosen, the $4\times4$ surface Hamiltonian (Eq.(\ref{eq:surfham})) is block diagonal leading to a doubly degenerate Dirac-like energy spectrum given by 
\begin{equation}
 E_{\bm{k},\pm}(\theta) = \pm\sqrt{v_{\|}^2k_y^2+v_1^2k_1^2}.
 \label{eq:spectrum}
\end{equation}
Since each surface of a strong 3D TI has only a single Dirac-like cone, the surface states are correctly described by taking linear combinations of two degenerate eigenstates of the surface Hamiltonian (Eq.(\ref{eq:surfham})). Such surface states can be written as 
\begin{equation}
 \Psi_{c(v)}^{1(2)}(\bm{k}) =\frac{1}{\sqrt{2}}\left(\psi_{E+(E-)}^+(\bm{k}) + e^{i \phi_R^{1(2)}}\psi_{E+(E-)}^-(\bm{k})\right) c^{1(2)}_{\bm{k},c(v)},
 \label{eq:surface}
\end{equation}
where $\psi^{+}_E(\bm{k})$ and $\psi^{-}_E(\bm{k})$ are the two states corresponding to energy $E$ and momentum $\bm{k}$ from the two degenerate eigenspaces of the surface Hamiltonian (Eq.(\ref{eq:surfham})), $1(2)$ in the superscript denotes top (bottom) surface, $c(v)$ in the subscript denotes conduction (valence) band, and $c^{1(2)}_{\bm{k},c(v)}$ is the corresponding electron annihilation operator with momentum {$\bm k$} for the TI surface. The phases $\phi_R^{1(2)}$ are fixed by ensuring the states in Eq.(\ref{eq:surface}) to be the positive (negative) eigenstates of the $T_x$ operator. From here onwards, we work only with the surface states corresponding to the positive eigenvalue of $T_x$ with the Fermi level placed in the conduction band ({\it i.e.} $\Psi^1_c(\bm{k})$).

Using the surface wavefunction given by Eq.~(\ref{eq:surface}) for an arbitrary surface termination characterized by the angle $\theta$ , the spin texture in the momentum space is obtained by taking the expectation of the spin operators $\mathds{I}_{2\times2}\otimes\bm{\sigma}$ as
\begin{equation}
\langle\bm\sigma(\bm{k})\rangle=\langle\Psi(\bm{k})\vert\mathds{I}_{2\times2}\otimes\bm{\sigma}\vert\Psi(\bm{k})\rangle,
\label{eq:spin_vector}
\end{equation}
which yields the expressions given in Eq.(\ref{eq:spinexpectation}) in the Appendix.

Note that the magnetization of the Fermi surface remains zero irrespective of non-triviality of the spin texture owing to time reversal symmetry. At the same time the magnetization of half of the Fermi surface carries unique signatures of the surface state corresponding to given $\theta$. This fact plays an central role in formulation of our strategy for identification of $\theta$ for a given arbitrary surfaces by using spin-polarized STM scans.


\section{Breaking of $\tau_z$~symmetry by surface states}\label{sec:degeneracy}
We first note that the block diagonal form of the $\mathcal{H}_{\text{surf}}(\theta)$ in Eq.~(\ref{eq:surfham}) implies that $\tau_z$ is a conserved quantity 
($i.e.,~[\mathcal{H}_{\text{surf}}(\theta),\tau_z\otimes \mathds{I}_{\sigma}]=0$) for the surface Hamiltonian. At the same time the surface states also have a two fold degeneracy 
where the degenerate solutions are given in Eq.~(\ref{eq:eigenstates4}). Though each of these individual degenerate solutions do respect the $\tau_z$ symmetry, a linear combination of 
these solutions  which is also a valid eigenstate of $\mathcal{H}_{\text{surf}}(\theta)$ could break the $\tau_z$ symmetry spontaneously. This is indeed the case for the surface 
states and this is a direct consequence of the fact that the surface states have to be simultaneous eigenstates of $\mathcal{H}_{\text{surf}}(\theta)$ and $ T_x~(=\alpha\, \tau_x\otimes\mathds{I}_{\sigma}+ \beta\, \tau_y\otimes\mathds{I}_{\sigma}$). As the operator $T_x$ lies on the $x$-$y$ plane in the $\tau$ space, the correct surface state are constructed as equal weight superposition of degenerate eigenstates of  $\tau_z\otimes\mathds{I}_{\sigma}$ with only a relative phase allowed between them (see Eq.~(\ref{eq:surface})) hence breaking the $\tau_z$ symmetry of $\mathcal{H}_{\text{surf}}(\theta)$.

\section{Perturbative calculation of tunneling current from STM}\label{sec:stm}
In this section, we engineer a minimal model for a tunnel Hamiltonian which can be used to inject spin-polarized electrons from a magnetic STM tip into the TI surface. The key ingredient in this construction stems from the observation that $\langle\tau_z\rangle$ is identically zero for all surface states for all momenta. The diagonal blocks of $\mathcal{H}_{\text{surf}}(\theta)$ (see Eq.~(\ref{eq:hammatrix}) of the Appendix) belong to the $+1$ or $-1$ eigenvalues of $\tau_z$ and have a spin degree of freedom associated with them (which points along $\bm{B}^+_{\bm{k}}$ and $\bm{B}^-_{\bm{k}}$ respectively as shown in the Appendix). Hence each STM electron injected into the TI surface state sees two independent channels given by  $\tau_z=\pm 1$. Tunneling into each channel corresponding to $\tau_z=\pm1$ will have a finite amplitude (call it $t^{\tau_{z}}_+$ and $t^{\tau_{z}}_-$). We note that the corresponding tunneling currents injected into each $\tau_z$ should not only be proportional to $|t_{\pm}^{\tau_z}|^2$ but also to the modulus squared of overlap of the STM electron spinor pointing along the polarization direction of the tip and that of a spinor which is pointing in the direction of $\bm{B}^\pm_{\bm{k}}$ respectively. This could give rise to a finite TMR which is the central focus of this article. As mentioned in Sec.\ref{sec:degeneracy}, the surface states indeed break the $\tau_z$ symmetry by being in a superposition state in the parity Hilbert space, specifically an equal weight superposition of the eigenstates of $\tau_z$, resulting in $\langle\tau_z\rangle=0$ for states corresponding to all $\theta$ and $\bm{k}$. Physically, the eigenstates of $\tau_z$ correspond to the two orbitals coming from the Bi and Se atoms, hence it is interesting to note that, within the low energy theory, on an average, the electronic orbitals corresponding to the Bi and Se atoms are exposed with equal weight on all surfaces. In a realistic situation, the relative phase between the two couplings $t^{\tau_z}_{+}$ and $t^{\tau_z}_{-}$ and their relative weights would depend on the microscopic details of overlap of the surface state wavefunction and the tip wavefunction, and is expected to get randomized over several measurements. Owing to this, and the fact that both the orbitals have equal weight on the surface states, an equal weight averaging over the relative phase and relative weight of $t^{\tau_z}_{+}$ and $t^{\tau_z}_{-}$ is performed at the end of the calculation to obtain the results which pertain to the physical situation. An efficient way to implement the above consideration is to artificially expand the Hilbert space of the tip Hamiltonian to include a fictitious $\tau$ degree of freedom on the tip so that we have 
\begin{equation}
 \Psi_{\text{STM}} ={\frac{1}{N}} \int dk ~ \begin{pmatrix}t^{\tau_z}_{+} \\t^{\tau_z}_{-}\end{pmatrix}\otimes\begin{pmatrix}\cos(\theta_s/2)\\ \sin(\theta_s/2)e^{i\phi_s}\end{pmatrix} e^{ikr}d_{k,\uparrow},
 \label{eq:stmstate}
\end{equation}
where $N$ is a normalization constant and the angles $\theta_s$ and $\phi_s$ denote the polar and the azimuthal angles for the magnetization direction of the STM tip, respectively. The corresponding Hamiltonian for the STM tip is given by   
\begin{equation}
 \mathcal{H}_{\text{STM}}=\sum_k \varepsilon_k d_{k,\uparrow}^{\dagger}d_{k,\uparrow}.
 \label{eq:hamstm}
\end{equation}
The tunnel Hamiltonian between the tip and surface can be written as 
\begin{equation}
 \mathcal{H}_{\text{tunn}} = J~(\Psi_{\text{STM}}^{\dagger}(\bm{r}=0)\Psi_c(\bm{r}=0) + \text{h.c}),
  \label{eq:tunnel}
\end{equation}
which in the momentum space looks like
\begin{equation}
 \mathcal{H}_{\text{tunn}} = J\sum_{\bm{k},k'}(z_{\bm{k}} c_{\bm{k}}^{\dagger}d_{k',\uparrow} + \text{h.c}),
\end{equation}
where $J$ is the tunneling amplitude, $z_{\bm{k}}$ is the overlap of the four component spinors of the TI (Eq.~(\ref{eq:genpsi})) and the STM (Eq.~(\ref{eq:stmstate})).

The corresponding current operator is defined as\cite{bruus2004many}
\begin{equation}
 \hat{I}=e\frac{d\hat{N}_{\text{STM}}}{dt} = \frac{ie}{\hbar}[\mathcal{H},\hat{N}_{\text{STM}}],
\end{equation}
where $\mathcal{H}$ is the sum of the TI, STM and the tunnel Hamiltonians. To leading order in perturbation theory, the expectation value of the current is given by
\begin{equation}
 \langle I\rangle=\frac{i}{\hbar}\int_{-\infty}^0 dt' \langle g\vert\left[\mathcal{H}_{\text{tunn},I}(t'),\hat{I}_{I}(0)\right]\vert g\rangle,
\end{equation}
where $\vert g\rangle =\vert g\rangle_{\text{TI}} \otimes \vert g\rangle_{\text{STM}}$ is the product of ground state of the individual systems in a decoupled state where both of them are maintained in equilibrium at two different Fermi energies which are infinitesimally close to one another. The difference of Fermi energies plays the role of applied bias. The subscript $I$  denotes the fact  that the operators are in the interaction picture. Writing everything in the momentum space, it becomes clear that the expectation value for the current operator can be written as a sum of momentum resolved currents if evaluated to lowest order in the tunnel coupling $J$ given by:
\begin{equation}
 \langle I\rangle=\int \frac{d\bm{k}}{N_1}~ {\Bigg {[}}~\frac{e \, J^2}{\hbar^2}~\vert~z_{\bm{k}}\vert^2\int \frac{dk'}{N_2}~\chi_{\bm{k},k'}~{\Bigg{]}},
 \label{eq:cur1}
\end{equation}
where 
\begin{equation}
 \chi_{\bm{k},k'} = \int_{-\infty}^{\infty}dt' \, \text{Im}\, [~ \mathcal{G}_{\text{TI}}(\bm{k},0;\bm{k},t') ~\mathcal{G}_{\text{STM}}(k',0;k',t')~].
 \label{eq:chikk}
\end{equation}
$\mathcal{G}$ denotes the standard time ordered fermionic Green's functions. $N_1$ and $N_2$ are  normalization constants, which depend on the system sizes of the TI and the STM respectively. The time integral in Eq.~(\ref{eq:chikk}) leads to a delta function $\delta(\varepsilon_k'-E_{\bm{k},+})$, which ensures energy conservation. The Green's functions just give the difference of Fermi functions, and the $\delta$- function in energy ensures appropriate conditions over $k'$ in Eq.~(\ref{eq:cur1}) for the evaluation of the integral over $k'$, leading to a momentum resolved expression for current which reads as 
\begin{equation}
 \langle I \rangle(\bm{k}) = \frac{eJ^2}{\hbar^2N_2}\vert z_{\bm{k}}\vert^2\chi_{\bm{k}},
\end{equation}
where
\begin{equation}
\begin{split}
 \chi_{\bm{k}} &= \int_{-\infty}^{\infty} dk'~\chi_{\bm{k},k'}\\
	       &=\hbar\rho^{\text{STM}}(n_{\text{F}}(E_{\bm{k}},\mu_{\text{TI}},T_{\text{TI}})-n_{\text{F}}(E_{\bm{k}},\mu_{\text{STM}},T_{\text{STM}})),
\end{split}
\end{equation}
where $n_{\text{F}}$ denotes the Fermi functions, and $\rho^{\text{STM}} = (\hbar v_{\text{F,STM}})^{-1}$ is the constant DOS of the STM. Note that, the STM is modeled as a 1D electron gas with a parabolic spectrum, however since we limit ourselves to tiny bias windows, the spectrum can be safely linearized and the DOS can be treated independent of the energy.

The expression for the momentum resolved current can be rearranged to explicitly identify the contribution coming from the three different processes to the leading order as follows,                                                                                         
\begin{align}
I(\bm{k})&=&\frac{e J^2 \rho^{\text{STM}}}{\hbar N_2}\left ({|t^{\tau_z}_{+}|^2\over2} M_1(\bm{k})+{|t^{\tau_z}_{-}|^2\over2} M_2(\bm{k})\right. \nonumber\\
&&\left.-2 \text{Re} [{(t^{\tau_z}_{+})}^{\ast}{t^{\tau_z}_{-}} M_{12}(\bm{k})]\right)\nonumber\\
&&\times(n_{\text{F}}(E_{\bm{k}},\mu_{\text{TI}},T_{\text{TI}})-n_{\text{F}}(E_{\bm{k}},\mu_{\text{STM}},T_{\text{STM}})).
\label{eq:cur2}
\end{align}
The first and the second term with the coefficients $\vert {t^{\tau_z}_{+}}\vert^2$ and $\vert {t^{\tau_z}_{-}}\vert^2$ refer to the processes where the electron is injected into and taken back from the even and odd parity orbitals respectively, hence they come with their respective STM coupling strengths $\vert {t^{\tau_z}_{+}} \vert^2$ and $\vert  {t^{\tau_z}_{-}} \vert^2$. The third term on the other hand refers to the process where the electron is injected into an even parity orbital but taken out from the odd parity orbital. The explicit form of the terms $M_1(\bm{k})$, $M_2(\bm{k})$ and $M_{12}(\bm{k})$,
\begin{equation}
M_1(\bm{k}) = 1 + \cos\theta_s\cos\theta_{\bm{k}}+\sin\theta_s\sin\theta_{\bm{k}}\cos(\phi_s-\phi_{\bm{k}})
\label{eq:M1}
\end{equation}
\begin{equation}
M_2(\bm{k}) = 1 - \cos\theta_s\cos\theta_{\bm{k}}+\sin\theta_s\sin\theta_{\bm{k}}\cos(\phi_s-\phi_{\bm{k}})
\label{eq:M2}
\end{equation}
\begin{align}
M_{12}(\bm{k})&=&e^{-i\phi_R}[\sin\theta_{\bm{k}}+\sin\theta_s ( \cos(\phi_s-\phi_{\bm{k}})+ \nonumber\\
&&\cos\theta_{\bm{k}}\sin(\phi_s-\phi_{\bm{k}}))]
\label{eq:M12}
\end{align}
sheds more light onto the three processes described above.  It is quite clear from Eq.~(\ref{eq:M1}) and Eq.~(\ref{eq:M2}) that $M_1(\bm{k})$ and $M_ 2(\bm{k})$ are nothing but the magnitude squared of the spinor overlaps of the spin part of the STM spinor and the ones representing the spins of the even and odd parity orbital sectors respectively. These two terms are later used to reconstruct the spin texture. The third term containing $M_{12}(\bm{k})$ can be understood in terms of a interference term for a two path interferometer where even and odd parity orbital sectors define the two paths. As discussed in the beginning of section, to obtain a physical answer we need to average over the relative amplitude and phase of $t_{\pm}^{\tau_z}$. In order to do so, we use the parameterization, $t^{\tau_z}_{+}= \cos(\theta_o/2)$ and $t^{\tau_z}_{-} =  e^{i\phi_o}\sin(\theta_o/2)$ where smallness of the tunneling amplitude is dumped into $J$ (see Eq.(\ref{eq:tunnel})). This  allows us to explore and perform an equal weight averaging over the full space of relative amplitudes of $t^{\tau_z}_{+}$ and $t^{\tau_z}_{-}$ and also an averaging over all possible relative phases (for $\theta_o$, averaging is done from $0$ to $\pi$, and for $\phi_o$,  it is  from $0$ to $2\pi$). The momentum resolved current before averaging takes a form given by 
\begin{eqnarray}
I(\bm{k})&=&\frac{e J^2 \rho^{\text{STM}}}{2\hbar N_2}\left[1 + \cos\theta_o\cos\theta_s\cos\theta_{\bm{k}} +\sin\theta_s \sin\theta_{\bm{k}} 
\right.\nonumber\\ 
&&\cos(\phi_s-\phi_{\bm{k}}) -\sin\theta_o\cos(\phi_o-\phi_R)\{\sin\theta_{\bm{k}}\nonumber\\
&& +\sin\theta_s(\cos(\phi_s-\phi_{\bm{k}}) +\left.\cos\theta_{\bm{k}}\sin(\phi_s-\phi_{\bm{k}}))\}\right] \nonumber\\
&&\times (n_{\text{F}}(E_{\bm{k}},\mu_{\text{TI}},T_{\text{TI}})-n_{\text{F}}(E_{\bm{k}},\mu_{\text{STM}},T_{\text{STM}}))
\label{eq:curk}
\end{eqnarray}
If we consider $\text{Bi}_2\text{Se}_3$, then the positive and negative parity orbitals approximately correspond to the orbitals of Bi and Se atoms respectively, hence surfaces with different orientation exposes the two orbitals with different weights due to arrangement of the Bi and Se atoms in the quintuple layer. Still an equal weight averaging over the strengths of tunneling ($t^{t_z}_{\pm}$)  by averaging over $\theta_o$ which quantifies the relative weight between orbitals of Bi and Se atoms, remains  justified in our analysis for the following reason. The terms which carry the information regarding this relative weight in Eq.\ref{eq:curk} are the those which depend on $\theta_o$. The terms involving $\sin\theta_o$ in Eq.(\ref{eq:curk}) always go to zero due to the averaging over $\phi_o$.  The other term involving $\cos\theta_o$ is multiplied to $\cos\theta_s$,  and since throughout our protocol for reconstruction of spin texture for surface state,  the STM magnetization is restricted to the global $x$-$y$ plane ($\theta_s=\pi/2$), the contribution of this term is also zero.  The point to note here is, as the average spin polarization of the momentum modes belonging to any surface state always stays in the $x$-$y$ plane (see Eq.(\ref{eq:spinexpectation})), hence the protocols used to reconstruct the information regarding the spin textures of the surface states does not require us to have an STM magnetization which has components along the global $z$ direction.  
After averaging over $\theta_o$ and $\phi_o$,  the momentum resolved current can be expressed as 
\begin{eqnarray}I(\bm{k})=& \frac{e J^2 \rho^{\text{STM}}}{2\hbar N_2} ( 1+ \sin \theta_s \sin \theta_{\bm{k}} \cos (\phi_s-\phi_{\bm{k}}))\times\nonumber\\ 
& (n_{\text{F}}(E_{\bm{k}},\mu_{\text{TI}},T_{\text{TI}})-n_{\text{F}}(E_{\bm{k}},\mu_{\text{STM}},T_{\text{STM}})).
          \end{eqnarray}
Since in an realistic situation, the STM will never be fully polarized, one has to account for it by putting in a polarization factor defined by $p=(\rho^{\text{STM}}_\uparrow - \rho^{\text{STM}}_\downarrow)/(\rho^{\text{STM}}_\uparrow + \rho^{\text{STM}}_\downarrow)$ in the expression for current as 
\begin{eqnarray}                 
 I(\bm{k})=&& \frac{e J^2 \rho^{\text{STM}}}{\hbar N_2}(1+p~\bm{S}_{\text{STM}}\cdot\langle\bm{\sigma}\rangle(\bm{k}))\times \nonumber\\
                  &&(n_{\text{F}}(E_{\bm{k}},\mu_{\text{TI}},T_{\text{TI}}) -  n_{\text{F}}(E_{\bm{k}},\mu_{\text{STM}},T_{\text{STM}})),
                           \label{eq:Ik}
                           \end{eqnarray}
where $\bm{S}_{\text{STM}}$ is the unit vector which is pointing along the magnetization of the STM. $\rho^{\text{STM}}=(\rho_\uparrow^{\text{STM}}+\rho_\downarrow^{\text{STM}})/2$ is defined as the average DOS of the STM. Note that the current injected from the STM into each momentum mode indeed has a elegant TMR form \cite{Slonczewski1989} once averaging is performed. Of course the contribution of the term $\bm{S}_{\text{STM}}\cdot\langle\bm{\sigma}\rangle(\bm{k})$ to $  I(\bm{k})$ reduces to zero if we sum over all momentum modes due to time reversal symmetry. On the other hand the value of  $\bm{S}_{\text{STM}}\cdot\langle\bm{\sigma}\rangle(\bm{k})$  when summed over a finite segment of the Fermi surface of the TI surface is non zero. As the total current turns out to be a momentum sum over the momentum resolved current (Eq.(\ref{eq:Ik})), one can define a current asymmetry which successfully captures the TMR response and can be measured in a multi-terminal set-up shown in Fig.~(\ref{fig:schematic}) by coupling to the magnetization of half of the Fermi surface. The current asymmetry ($\Delta I$) is defined as
\begin{equation}
\Delta I  = I_{L}-I_{R}= \left(\int\limits_{\gamma}^{\gamma+\pi} - \int\limits_{\gamma+\pi}^{\gamma+2\pi}\right)d\delta_{\bm{k}}~\int\limits_0^\infty \frac{dk}{N_1} ~k~ I(\bm{k}),
\label{eq:deltai}
\end{equation}
where $k_1$ and $k_y$ are parameterized as $k_1 = k\cos\delta_{\bm{k}}$ and $k_y = k\sin\delta_{\bm{k}}$ and $I_L$ and $I_R$ are the current collected by the left and the right contact defined consistently as in Fig.(\ref{fig:schematic}). From Eqs.(\ref{eq:Ik}) and (\ref{eq:deltai}), it can be seen that 
\begin{equation}\Delta I \propto\left(\int\limits_{\gamma}^{\gamma+\pi} - \int\limits_{\gamma+\pi}^{\gamma+2\pi}\right)d\delta_{\bm{k}}~\int\limits_0^\infty dk~ \bm{S}_{\text{STM}}\cdot\langle\mathbf{\sigma}\rangle(\mathbf{k}).
\end{equation} Owing to the presence of time reversal symmetry, 
\begin{equation}
\int\limits_{\gamma}^{\gamma+\pi} d\delta_{\bm{k}}~\int\limits_0^\infty dk~ \langle\mathbf{\sigma}\rangle(\mathbf{k})=- \int\limits_{\gamma+\pi}^{\gamma+2\pi}d\delta_{\bm{k}}~\int\limits_0^\infty dk~ \langle\mathbf{\sigma}\rangle(\mathbf{k}),
\end{equation}
which means that $\Delta I$ essentially couples to the magnetization of half of the Fermi surface of the TI surface state where the chosen half of the Fermi surface depends on the choice of $\gamma$.
 This result gives a clear indication that, by studying the profile of the current asymmetry $\Delta I $ as a function of the  magnetization direction of the STM electron and $\gamma$, one can reconstruct the Fermi surface and  its spin texture for all surfaces corresponding to different values of $\theta$.  To get rid of the dependencies on parameters like the system sizes, tunneling strength etc., one can always define a dimensionless measurable quantity which is the ratio of the current asymmetry $\Delta I$ and the  total injected current  $I_0$ given by
\begin{equation}
\frac{\Delta I}{I_0}= \frac{ \left(\int\limits_{\gamma}^{\gamma+\pi} - \int\limits_{\gamma+\pi}^{\gamma+2\pi}\right)d\delta_{\bm{k}}~\int\limits_0^\infty dk ~k ~I(\bm{k})}{ \int\limits_{0}^{2\pi}d\delta_{\bm{k}}~\int\limits_0^\infty dk ~k~ I(\bm{k})}.
\end{equation}
\section{Reconstruction of spin texture}\label{sec:spintexture}
The objective of this section is to show how the current asymmetry measurements lead to a direct reconstruction of the spin textures of Fermi surfaces of arbitrary TI surface states. It was earlier shown in Ref.~[\onlinecite{Roy2014}] that scanning $\Delta I$ as a function of $\gamma$ with a fixed chosen direction for magnetization of the STM tip for a perfectly spin-momentum locked $\theta=0$ surface leads to a complete  reconstruction of the spin texture on the Fermi surface. This was possible as the Fermi surface for the $\theta =0$ surface has a perfect azimuthal symmetry in the momentum space.  For such a circular Fermi surface, the relative angle between the STM magnetization, and $\gamma$ for which the TMR signal is extremum gives the spin-momentum locking angle and the sign of the extremum gives the chirality, thus completely reconstructing the spin-texture. However, for an arbitrary surface ($\theta\ne0$), the Fermi surface is elliptical as can be seen from Fig.~(\ref{fig:spectrum}) in the Appendix. Also, the average spin associated with each momentum mode is not a constant as we move along the Fermi surface. Moreover, the spin-momentum locking angle is also skewed in the sense that the angle between the averaged spin and the corresponding momentum is not a constant as we move along the Fermi surface. The Fermi surface being elliptic and the spin momentum locking angle being not a constant  make it necessary that we make a TMR scan over $\gamma$ for at least two magnetizations directions of the STM spin for a unique reconstruction of the spin texture. This is so because, for a general $\theta$, the $x$- and $y$-components of the spins have independent dynamics as we move along the Fermi surface (as evident from Eq.(\ref{eq:spinexpectation}) in the Appendix) and hence it requires two distinct measurements with different STM magnetization to reconstruct them . 

The formal expression (Eq.~(\ref{eq:Ik})) governing the TMR response is same for all surfaces irrespective of the details of the spin texture discussed above (though the actual response depends on the details of the spin texture) and this is attributed to the fact that the current asymmetry $\Delta I$ couples to the magnetization density (per unit area) of the half of the Fermi surface (i.e., $\langle\bm{\sigma}\rangle_{\text{half}}$) in the bias window  which is defined as
\begin{equation}
\langle\bm{\sigma}\rangle_{\text{half}}(\gamma) = \int_0^{\infty}kdk \int_{\gamma}^{\gamma+\pi}d\delta_{\bm{k}} \langle\bm{\sigma}\rangle(n_{\text{F,TI}}-n_{\text{F,STM}}).
\end{equation}
Presence of time reversal symmetry forces the net magnetization over the full Fermi surface to be zero, which implies that the magnetization of the two halves of the Fermi surface are equal and opposite to each other, i.e $\langle\bm{\sigma}\rangle_{\text{half}}(\gamma)=-\langle\bm{\sigma}\rangle_{\text{half}}(\gamma+\pi)$. Since the current asymmetry basically couples to the difference in the magnetization of the two halves i.e $\Delta I \sim \bf{S}_\text{STM}\cdot(\langle\bm{\sigma}\rangle_{\text{half}}(\gamma)-\langle\bm{\sigma}\rangle_{\text{half}}(\gamma+\pi))$, it is easy to see that 
\begin{equation}
\Delta I  \sim  \bf{S}_\text{STM}\cdot\langle\bm{\sigma}\rangle_{\text{half}}(\gamma),
\end{equation}
which is the central relation connecting the current asymmetry to the spin texture and hence leading to its reconstruction. The magnetization of half of the Fermi surface, depends on two factors, $\theta$ and $\gamma$. These two factors uniquely fix the magnetization of the given segment of the Fermi surface. 
\begin{figure}
\begin{center}
\includegraphics[width=0.99\columnwidth]{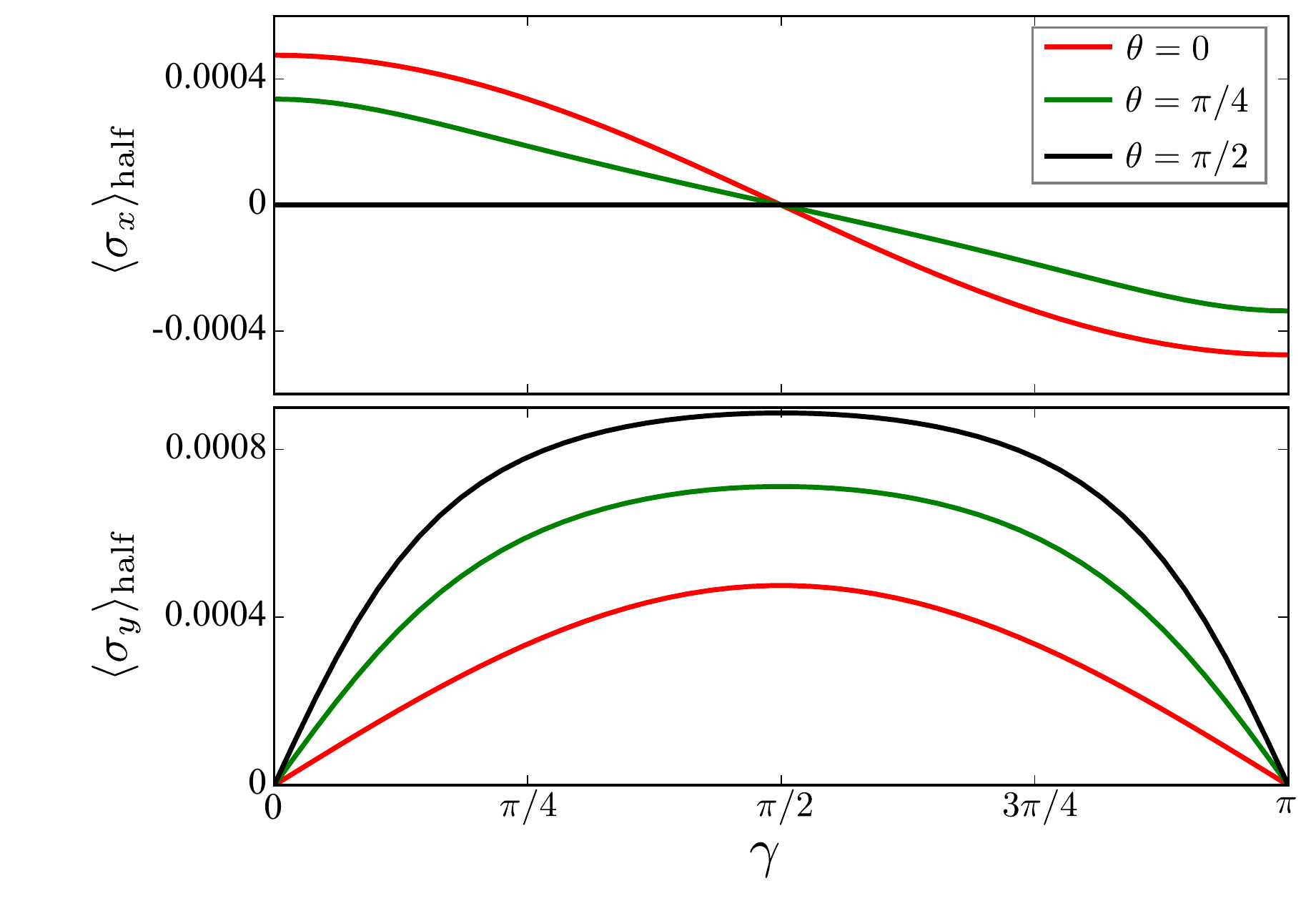}
\end{center}
\caption{The panels show the magnetization of the half of the Fermi surface in the $x$ direction (top) and the $y$ direction (bottom) as a function of $\gamma$ for three different surfaces corresponding to $\theta=0, \pi/4 \text{ and }\pi/2.$}
\label{fig:spinhalf}
\end{figure}
The strategy which is followed for the spin reconstruction is as follows: first, a TMR scan by varying $\gamma$ is done for an STM magnetized in the $x$ direction and the current asymmetry is plotted as a function of $\gamma$.  The sign of the asymmetry at $\gamma=0$ provides the handedness of the spin texture. For reconstructing the magnitude of the spin another TMR scan needs to be done by varying  $\gamma$  but now with the STM magnetization pointing in the $y$ direction. 
\begin{figure}
\begin{center}
\includegraphics[width=0.99\columnwidth]{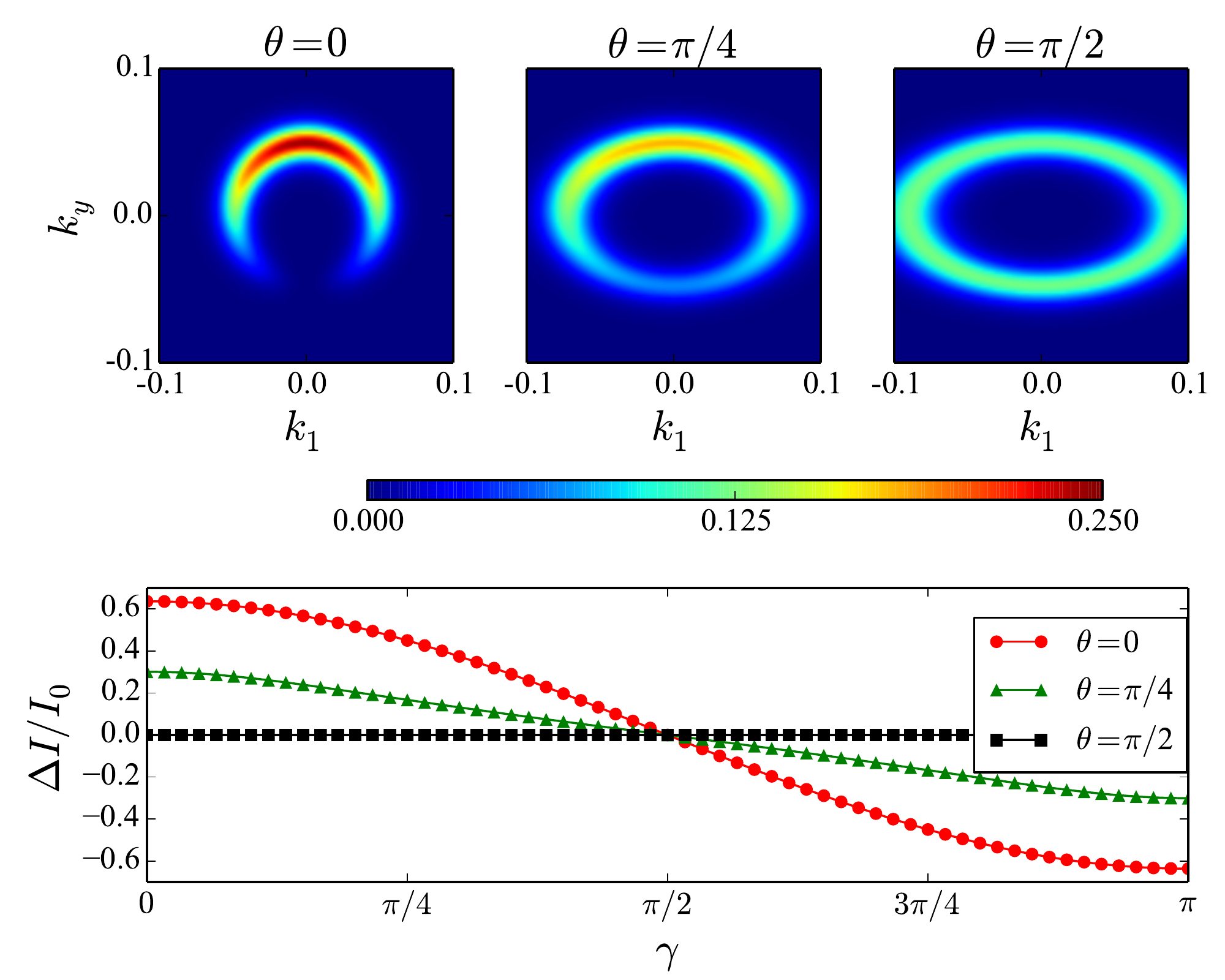}
\end{center}
\caption{Top: The three panels show the current distribution in the $k_1-k_y$ plane (in units of $\text{\AA}^{-1}$) for three different surfaces with the STM spin-polarized in the $x$-direction. Bottom: The TMR response corresponding to the same situation is shown by plotting $\Delta I /I_0$ as a function of $\gamma$.}
\label{fig:curasymspinx}
\end{figure}
$\Delta I/I_0$ plotted for the case of STM spin pointing in the $x$-direction is shown in Fig.~(\ref{fig:curasymspinx}). The three panels at the top show the current distribution pattern at finite bias  in the $k_1-k_y$ plane for  three different surfaces $\Sigma(\theta)$ corresponding to $\theta=0,\pi/4$ and $\pi/2$ as labeled in the plot. It can be seen that,  as one moves from the $\theta=0$ surface to an oblique surface, the anisotropy in current distribution (Fig.~(\ref{fig:curasymspinx}) top) and the resulting experimentally measurable current asymmetry (Fig.~(\ref{fig:curasymspinx}) bottom) decreases and eventually goes to $0$ on the side surfaces. The current asymmetry for this  case is a direct reflection of the spin polarization in the $x$ direction, as it exactly mirrors the $x$-component of  magnetization of half the Fermi surface as shown in Fig.~(\ref{fig:spinhalf}). Hence such a measurement will indeed lead to reconstruction of  $x$-component  of the spin texture presented in  Fig.~(\ref{fig:spinhalf}).

\begin{figure}
\begin{center}
\includegraphics[width=0.99\columnwidth]{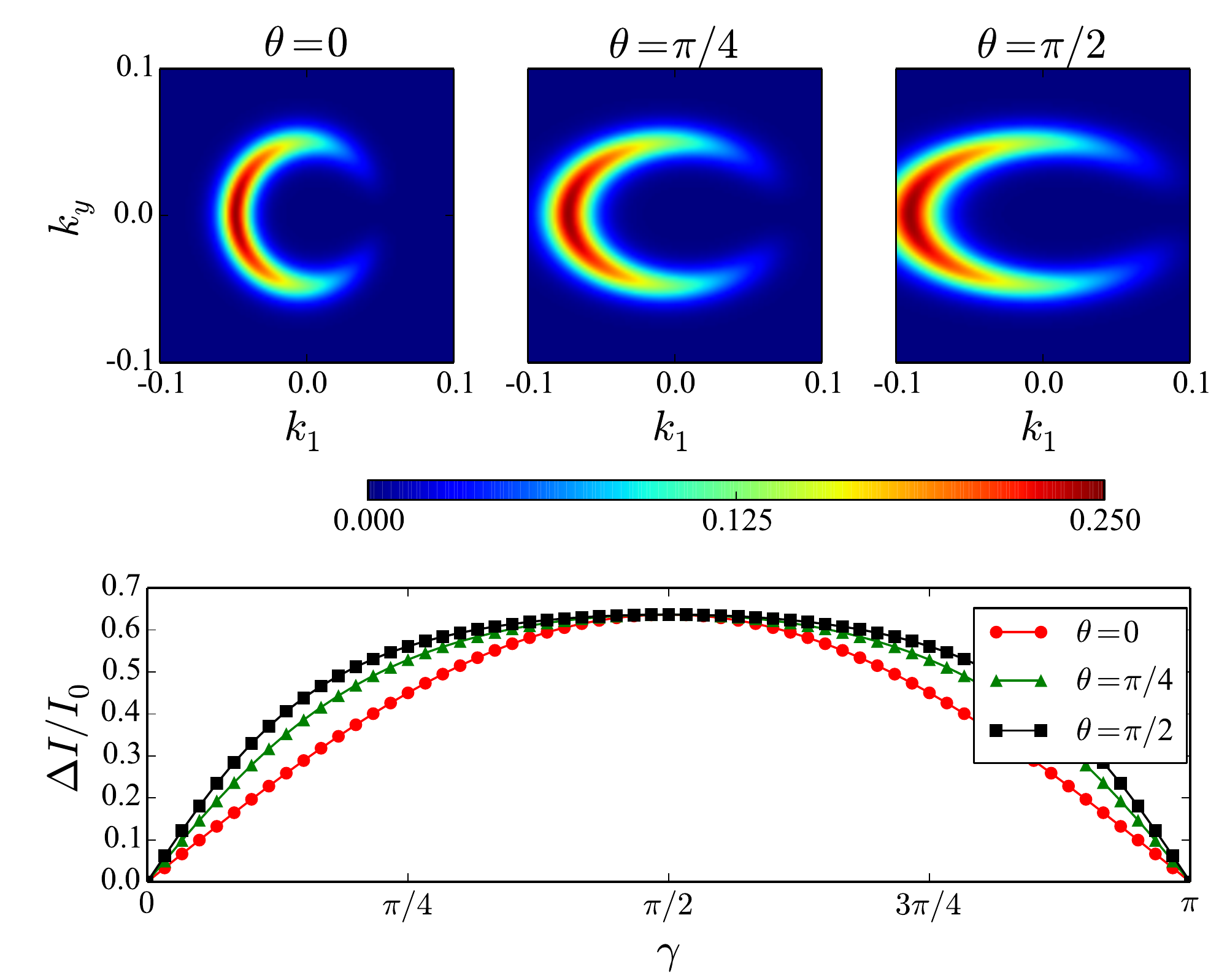}
\end{center}
\caption{Top: The three panels show the current distribution in the $k_1-k_y$ plane (in units of $\text{\AA}^{-1}$) for three different surfaces with the STM spin-polarized in the $y$ direction. Bottom: The TMR response corresponding to the same situation is shown by plotting $\Delta I /I_0$ as a function of $\gamma$.}
\label{fig:curasymspiny}
\end{figure}

These measurements however do not throw any light on the $y$-magnetization of the Fermi surface and hence one can not predict anything about the texture of the magnitude of the spin polarization of the surface state based on the above measurement alone. For this one needs, as mentioned above, another similar set of measurements, but with the STM spin pointing in the $y$-direction. These results are presented in Fig.~(\ref{fig:curasymspiny}).
The current asymmetry peak as a function of $\gamma$ for the STM polarized in the $y$ direction (Fig.~(\ref{fig:curasymspiny}) bottom) is shifted by $\pi/2$ compared to the case where the STM polarization is in the $x$ direction (Fig.~(\ref{fig:curasymspinx}) bottom). However, for oblique surfaces the asymmetry profile is not just a trivial shift of the curve by $\pi/2$. Two differences are to be noted here, firstly, the asymmetry does not go down as one goes to a more and more oblique surface and secondly, the curve flattens out more and more near the peak  as one goes to a more and more oblique surface. The first point implies that magnitude of the spin polarization does not go down as $\theta$ is increased, and the second, points towards an elliptical Fermi surface. This can be understood as follows: from Eq.~(\ref{eq:sy}) one can see that on the Fermi surface, $\langle\sigma_y\rangle(\bm{k})$ is directly proportional to $k_1$. For a wide range of angles near the peak, $k_1$ is high enough to flatten out the peak of the current asymmetry. This already suggests an extended structure of the Fermi surface in the $k_1$ direction. Also observation that as $\theta$ is increased the fall of the asymmetry's magnitude to $0$ gets steeper. This also indicates that the Fermi surface is less extended in the $k_y$ direction compared to the $k_1$ direction. These two observations uniquely identify the conic section of the Fermi surface to be an ellipse. Another information that can be extracted is about the Fermi velocities perpendicular and parallel to the crystal growth axis ($v_{\|}$ and $v_z$ respectively). As one goes to a more and more oblique surface, one expects that the contribution of $v_z$ to the Fermi velocity in $k_1$ direction to increase. So the observation that as $\theta$ is increased the Fermi surface gets more elliptical and extended in the $k_1$ direction points to the fact that $v_z<v_{\|}$ as the semi-axes of the Fermi surface in the $k_1$ and $k_y$ directions are proportional to $v_1^{-1}$ and $v_{\|}^{-1}$ respectively. Coming back to the spin texture, one can conclude that all the surfaces are devoid of any spin polarization in the $z$ direction as one never finds any current asymmetry, if the STM is spin-polarized in the $z$ direction. Hence in this section we have shown, how the current asymmetry can uniquely reconstruct the spin texture $\langle\sigma_x\rangle(\bm{k})$ and $\langle\sigma_y\rangle(\bm{k})$ and  provide clear indication of change of Fermi surface from a circular (for $\theta =0$ surface) shape to an ellipse (for $\theta \neq 0$ surface) like shape.\\


\section{Spin texture as hologram of the bulk}\label{sec:hologram}
In this section, we show that the spin texture reconstruction via current asymmetry measurement $\Delta I$ could lead to a unique identification of the surface $\Sigma(\theta)$ and the $\theta$-dependent Fermi velocities of the surface states. These in turn can uniquely determine $v_z$ and $v_{\|}$, thus giving information of the bulk band structure of the crystal. The strategy for this identification is based on the observation that the  expression for the $\langle\sigma_x\rangle$ in Eq.~(\ref{eq:sx}) has a momentum independent part which is proportional to $\cos\theta/v_3$. This observation implies that  the $x$-component of magnetization for any finite segment of the Fermi surface also scales by this factor which depends only on the value of $\theta$. Hence this factor will show up in the expression for  $\Delta I$ (note that $\Delta I$ is proportional to magnetization of half of the Fermi surface) provided we choose the STM magnetization along the $x$-direction. To see this fact more clearly we note that the momentum resolved tunneling current for the case of STM magnetized pointing along $x$ direction (obtained from Eq.~(\ref{eq:Ik})) is given by
\begin{equation}
I(\bm{k}) \propto (1+p~\langle\sigma_x\rangle(\bm{k})).
\end{equation}
It is now obvious from Eq.~(\ref{eq:deltai}) that the current asymmetry $\Delta I$ in this situation has to be proportional to the $\langle\sigma_x\rangle$ averaged over half of the Fermi surface of the surface state, which in turn depends on the scale factor $\cos\theta/v_3$. To maximize the signal, orientation of the contact is taken to be such that it maximizes $\Delta I$ for the given magnetization direction of the STM tip. It can be shown that $\gamma=0$ is the configuration\cite{Roy2014} that meets the criterion. Note that, changing the orientation of the contact (i.e., changing  $\gamma$) on the surface must be done with respect to the  orientation of the underlying lattice which is be to be held fixed. Hence changing the orientation of  contact is not equivalent to rotating the sample about the $k_3$-axis. \\

\begin{figure}[t]
\begin{center}
\includegraphics[width=0.9\columnwidth]{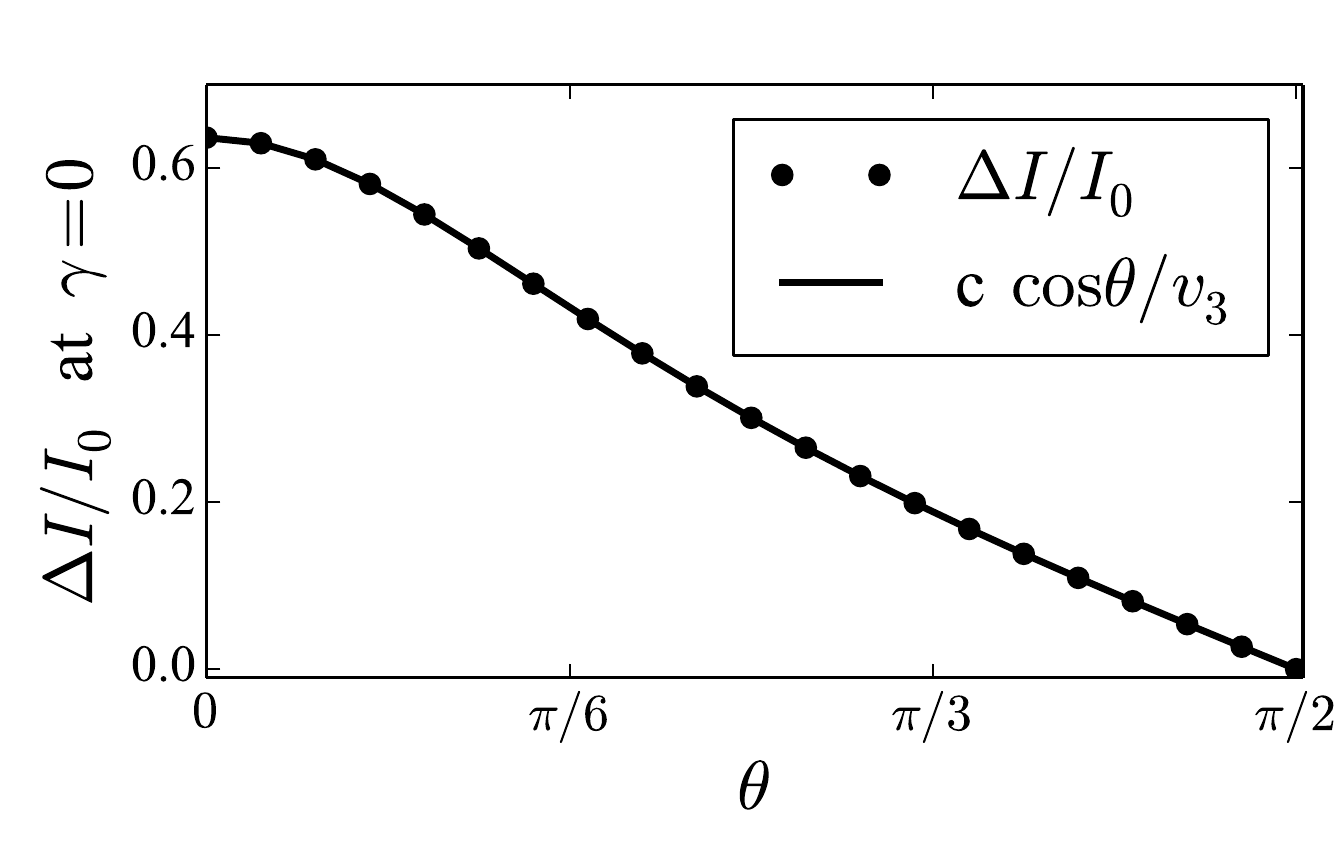}
\end{center}
\caption{The data points show $\Delta I/I_0$ calculated for the STM tip being magnetized in the $x$ direction ($\theta_s=\pi/2,\phi_s=0$) and the ideal case $p=1$, for the contact configuration corresponding to $\gamma=0$. The solid line shows the continuous function $\text{c}~\cos\theta/v_3$ which is the function followed by the current asymmetry as function of the surface.}
\label{fig:surface}
\end{figure}

A systematic measurement of the quantity $\Delta I$ as a function of $\theta$ while the STM magnetization points only along the $x$-direction should fit a function of the form $c\,( \cos\theta/v_3) $ where $c$ is a constant. Hence, for an arbitrary surface $\langle\sigma_x\rangle$, measurement of $\Delta I$ with the STM magnetization pointing along $x$-direction can be used to uniquely identify the value of $\theta$ for that particular surface. Also, to ensure that a possible variation in strength of the tunnel coupling $J$ used to probe surfaces with different $\theta$ does not spoil the proposed strategy for identifying the surface, one can always normalize $\Delta I$ by total injected current $I_0=I_R+I_L$ resulting in a dimensionless quantity irrespective of the details of $J$. This fact is demonstrated in Fig.~(\ref{fig:surface}). The value $c$ appearing in the plot is given by the following relation,
 \begin{equation}
 c = v_z \left.\frac{\Delta I}{I_0}\right\vert_{\theta=0,\gamma=0}.
 \label{eq:c1}
 \end{equation}
 Since, for the $\theta=0$ surface,  the spin momentum locking is perfect, it is straightforward to show that 
 \begin{equation}
 \left.\frac{\Delta I}{I_0}\right\vert_{\theta=0,\gamma=0} = p~\frac{2}{\pi}.
 \label{eq:c2}
 \end{equation}
 Plugging Eq.~(\ref{eq:c2}) in Eq.~(\ref{eq:c1}), one gets the value of $c$ as 
  \begin{equation}
 c = p~ 2v_z/\pi.
 \end{equation}

Now we observe that $\Delta I/I_0$ if experimentally measured could be fitted to a theoretically predicted functional form given by
\begin{equation}
\frac{\Delta I}{I_0} = p~\frac{2v_z}{\pi}\frac{\cos\theta}{\sqrt{v_z^2\cos^2\theta + v_{\|}^2\sin^2\theta}},
\end{equation}
which is just a two-parameter fit over the parameters $v_z$ and $v_{\|}$ of the bulk Hamiltonian. Hence this fit could lead to a read out of  $v_z$ and $v_{\|}$.  In conclusion, the collection of measurements of the current asymmetries over different surfaces, acts like a hologram of the bulk. Such a reconstructions of bulk parameters is impossible with any set of measurements on a single surface.

\section{Conclusion}\label{sec:conclusion}
In this article we have developed a tunnel Hamiltonian approach for injection of spin-polarized electrons from a magnetized STM into the planar  surface which is exposed by cleaving a crystal of a  3D TI  material like $\text{Bi}_2\text{Se}_3$ at an arbitrary angle $\theta$ with respect to the crystal growth axis.  We apply this tunnel Hamiltonian approach to demonstrate that, electrical transport in the linear response regime carries unique signatures of the surface corresponding to a given $\theta$.  Our proposal also provides a strategy to exploit the three terminal TMR scan to determine the unconventional spin-momentum locking textures for the  $\theta \neq 0,\pi$ surfaces. We show that a complete set of current asymmetry measurements over different surfaces can actually be used to identify the material parameters which characterize the bulk band structure  and hence acts as a hologram. The TMR scan can also identify the angle $\theta$ for an arbitrary surface, hence, though conventionally x-ray spectroscopy is used to identify different crystal surfaces, in this case one can use an electrical transport probe as an alternative. \\

\section*{Acknowledgments} The authors thank Diptiman Sen for discussions and Krishanu Roychowdhury for his critical reading of the manuscript. KS is funded by U. de Sherbrooke, Qu\'ebec's RQMP and Canada's NSERC.

\appendix

\section{Derivation of the surface states and their spin textures}\label{sec:app}
The appendix contains a self contained derivation of the expressions of the surface states and their spin textures as obtained in Ref.[\onlinecite{Zhang2012}].

Explicitly writing down the surface Hamiltonian in Eq.(\ref{eq:surfham}) in our basis reveals that it is already in a block diagonal form (which was the reason for choosing the $\bm{\tau}\otimes \bm{\sigma}$ representation) and now each block can be written in a $\bm{\sigma}.\bm{B}$ form where $\bm{B}$ can be thought of as an effective magnetic field acting on the spin degree of freedom belonging to the respective blocks. The surface Hamiltonian in Eq.(\ref{eq:surfham}) written in $\bm{\tau}\otimes \bm{\sigma}$ basis takes the block diagonal form given by 
\begin{equation}
 \mathcal{H}_{\text{surf}}(\theta)=\begin{pmatrix}
                                    \bm{\sigma}.\bm{B}^+_{\bm{k}}(\theta) && 0 \\
                                    0 &&\bm{\sigma}.\bm{B}^-_{\bm{k}}(\theta)
                                   \end{pmatrix},
\label{eq:hammatrix}
\end{equation}
where
\begin{equation}
\bm{B}^\pm_{\bm{k}}(\theta)=\{v_{\|}k_y\alpha,-v_1k_1,\mp v_{\|}k_y\beta\}.
\label{eq:bfield}
\end{equation}
\begin{figure}[t]
\begin{center}
\includegraphics[width=\columnwidth]{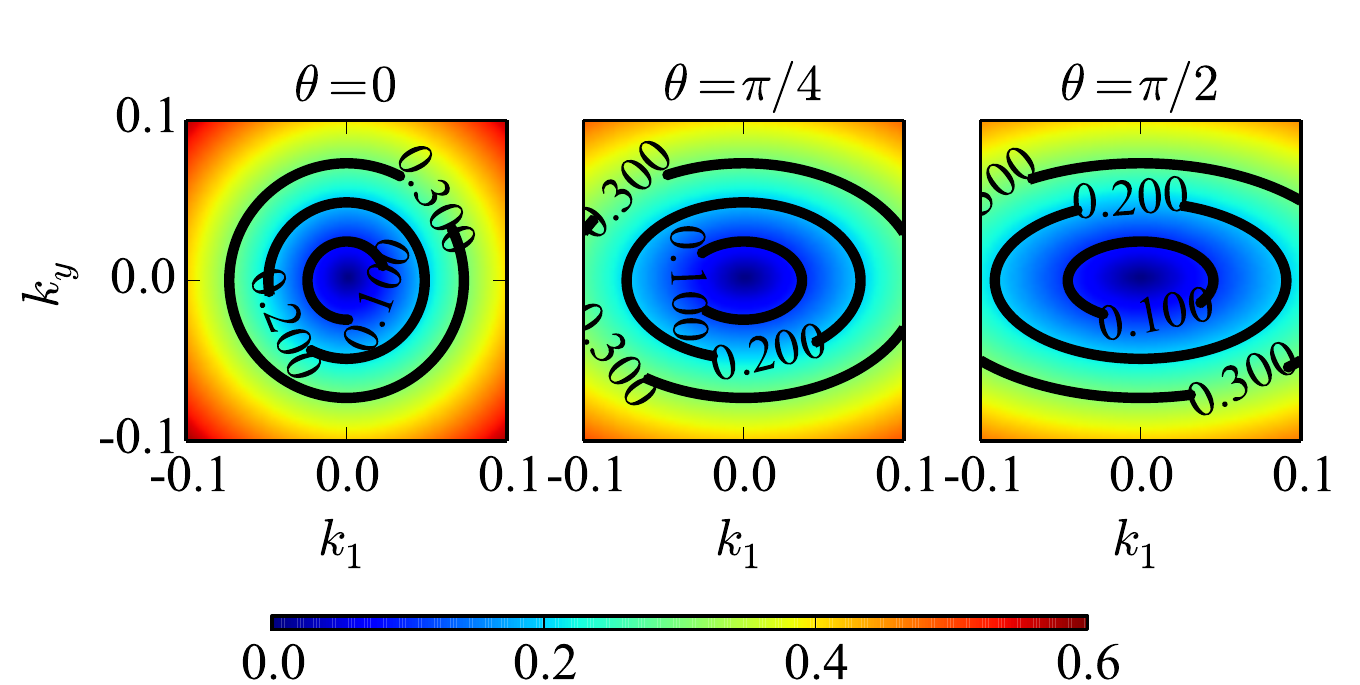}
\end{center}
\caption{The three panels show the energy spectrum of the conduction band for three different surfaces. The thick lines show the three constant energy contours corresponding to 0.1, 0.2 and 0.3 eV respectively. It should be noted that as one moves from the top surface to the side surface, the constant energy contour begins to go from a circle to a ellipse. This happens because the contribution of  $k_z$ begins to grow in the in-plane components of the momentum which magnifies the effect of the  difference of the two Fermi velocities $v_z$ and $v_{\|}$. The axes of the figures correspond to the in-plane momenta (in units of $\text{\AA}^{-1}$)}.
\label{fig:spectrum}
\end{figure}
Since $\vert\bm{B}_{\bm{k}}^+(\theta)\vert = \vert\bm{B}_{\bm{k}}^-(\theta)\vert$, the energy dispersion resulting from Eq.~(\ref{eq:hammatrix}) is two-fold degenerate. Also note that $\bm{B}_{\bm{k}}^+(\theta) \text{ and } \bm{B}_{\bm{k}}^-(\theta)$ differ only in their $z$-component with an equal magnitude and opposite sign. This fact  gives clear indications that, if we evaluate the expectation value of spin in an eigenstate of this Hamiltonian, its $z$-component may be zero regardless of the value of $\theta $ as will be evident in the next section.   The surface dependent energy dispersion in Eq.(\ref{eq:spectrum}) is plotted in Fig.(\ref{fig:spectrum}) for three different $\theta$s and it shows how the Fermi surface evolves from a circular one to an elliptical one as one goes from $\theta=0$ to $\theta=\pi/2$.
The surface Hamiltonian in Eq.~(\ref{eq:hammatrix}) is a $4\times4$ matrix, giving rise to the two degenerate Dirac-like dispersions, however it should be noted that one exposed surface of the TI has only one Dirac-like cone living on it. Actually one Dirac cone belongs to the top surface and the other one belongs to the bottom one. This suggests that an arbitrary state picked from the degenerate subspace of the eigenstates of $\mathcal{H}_{\text{surf}}(\theta)$ does not represent the true surfaces states.  The correct four component spinor describing the appropriate surfaces states are constructed in the following way. 

The two eigenstates for each of the two blocks in Eq.~(\ref{eq:hammatrix}) are given by
\begin{equation}
 \chi^{\pm}_{E+}(\bm{k}) = \begin{pmatrix}
                    \cos\frac{\theta^{\pm}_{\bm{k}}}{2}\\
                    \sin\frac{\theta^{\pm}_{\bm{k}}}{2}e^{i\phi_{\bm{k}}^{\pm}}
                   \end{pmatrix}; ~~
  \chi^{\pm}_{E-}(\bm{k}) = \begin{pmatrix}
                    -\sin\frac{\theta^{\pm}_{\bm{k}}}{2}\\
                    \cos\frac{\theta^{\pm}_{\bm{k}}}{2}e^{i\phi_{\bm{k}}^{\pm}}
                   \end{pmatrix},
\label{eq:eigenstates2}
\end{equation}
where $\chi^{+}_{E+}$ and $\chi^{+}_{E-}$ are the positive and negative energy eigenstates respectively of the first block, and $\chi^{-}_{E+}$ and $\chi^{-}_{E-}$ are of the second block of Eq.~(\ref{eq:hammatrix}). With this, it is straightforward to construct the eigenstates of the Hamiltonian in Eq.(\ref{eq:hammatrix}) as
\begin{eqnarray}
 \psi^+_{E+} = \begin{pmatrix}
                \chi^{+}_{E+}\\
                \bm{0}
               \end{pmatrix}; ~~&
  \psi^-_{E+} = \begin{pmatrix}
		\bm{0}\\
                \chi^{-}_{E+}
                \end{pmatrix}
\nonumber\\
 \psi^+_{E-} = \begin{pmatrix}
                \chi^{+}_{E-}\\
                \bm{0}
               \end{pmatrix}; ~~&
  \psi^-_{E-} = \begin{pmatrix}
		\bm{0}\\
                \chi^{-}_{E-} 
                \end{pmatrix} .
\label{eq:eigenstates4}
\end{eqnarray}
In Eqs.~(\ref{eq:eigenstates2}) and (\ref{eq:eigenstates4}), $\theta_{\bm{k}}^\pm$ and $\phi_{\bm{k}}^\pm$ are defined via the effective magnetic fields (Eq.~(\ref{eq:bfield})) as
\begin{equation}
 \bm{B}^\pm_{\bm{k}}=\vert\bm{B}_{\bm{k}}^\pm\vert(\sin\theta_{\bm{k}}^\pm\cos\phi_{\bm{k}}^\pm,\sin\theta_{\bm{k}}^\pm\sin\phi_{\bm{k}}^\pm,\cos\theta_{\bm{k}}^\pm).
\end{equation}
The relation between $\theta_{\bm{k}}^\pm$ and $\phi_{\bm{k}}^\pm$ can be trivially seen from Eq.~(\ref{eq:bfield}) as $\theta_{\bm{k}}^-=\pi-\theta_{\bm{k}}^+$ and $\phi_{\bm{k}}^-=\phi_{\bm{k}}^+$. From now on, we drop the $\pm$ superscript on the $\theta_{\bm{k}}^\pm$ and $\phi_{\bm{k}}^\pm$ and write everything in terms of $\theta_{\bm{k}}^+$ and $\phi_{\bm{k}}^+$ denoted by $\theta_{\bm{k}}$ and $\phi_{\bm{k}}$. 

The correct description of states on the surface of the topological insulator can now be constructed by taking appropriate linear combinations of the states from the two fold degenerate subspace discussed above.  The coefficients of the linear combination are fixed by considering the fact that the surface state also has to be an eigenstate of $T_x$ operator. The surface states being an eigenstate of $T_x$ operator can be understood from the last terms in the expression for bulk Hamiltonian $\mathcal{H}$ in Eq.~(\ref{eq:bulkhamlocal}) which finally induces the dispersing modes of the surface Hamiltonian. Under the above consideration, the wavefunction for surface electron (Eq.(\ref{eq:surface}) of the main text) can be written as
 
\begin{equation}
\Psi_{c(v)}^{1(2)}(\bm{k}) =\frac{1}{\sqrt{2}} \begin{pmatrix}
						\chi_{E+(E-)}^+(\bm{k})\\
						e^{i \phi_R^{1(2)}}\chi_{E+(E-)}^-(\bm{k})\\
					\end{pmatrix} c^{1(2)}_{\bm{k,c(v)}}, 	
\label{eq:genpsi}
\end{equation}

\begin{figure}[t]
\begin{center}
\includegraphics[width=0.95\columnwidth]{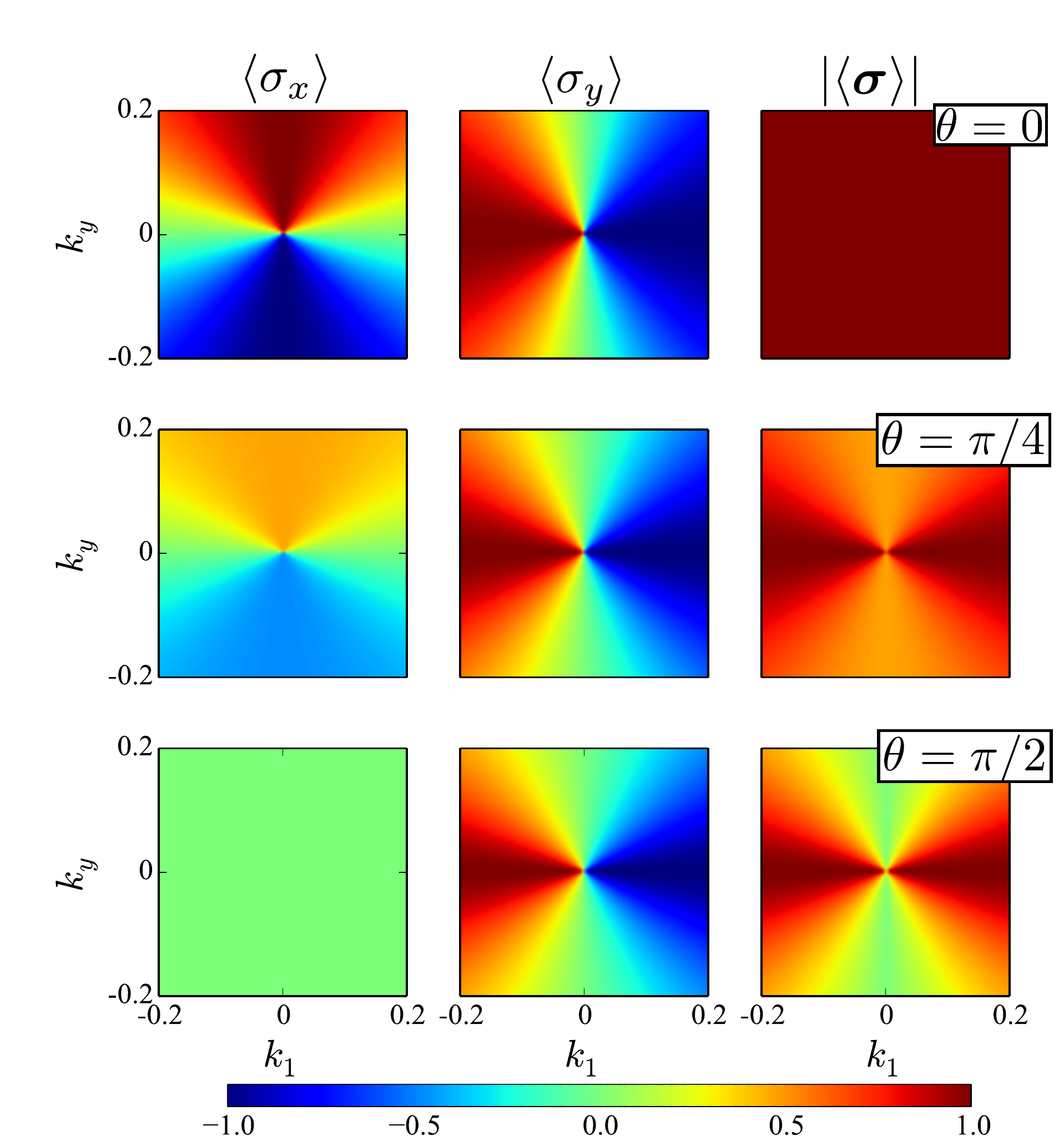}
\end{center}
\caption{The spin texture for three different surfaces corresponding to $\theta=0,\pi/4, \text{ and } \pi/2$ are plotted in the momentum space. It should be noted that the axes of the figures are $k_1$ and $k_y$ (in units of $\text{\AA}^{-1}$) which are the local in-plane axes, however the first two columns show the expectation of spin along the global $x$ and $y$ directions. Since $\langle\sigma_z\rangle$ is identically zero, it is not plotted, rather the expectation of the total spin magnitude is plotted to show that it is not constant on the Fermi surface for arbitrary surfaces.}
\label{fig:spintexture}
\end{figure}

The expectation value of the $\mathds{I}_{2\times2}\otimes\bm{\sigma}$ in the state  given by Eq.~(\ref{eq:genpsi}) yields the spin texture as 
\begin{equation}
\langle\Psi_c\vert\mathds{I}_{2\times2}\otimes\bm{\sigma}\vert\Psi_c\rangle =\frac{1}{2}( \langle\chi_{E+}^+\vert\bm{\sigma}\vert\chi_{E+}^+\rangle + \langle\chi_{E+}^-\vert\bm{\sigma}\vert\chi_{E+}^-\rangle).
\label{eq:spin_vector}
\end{equation}
Note that the effective magnetic field for the two parity sectors are related as $B_{x}^+=B_{x}^-$, $B_{y}^+=B_{y}^-$, and $B_{z}^+=-B_{z}^-$. This implies that $\langle\sigma^z\rangle$ should identically vanish for any arbitrary surface. The expectation value of the spin can be directly read off from the Hamiltonian as
\begin{equation}
 \langle\bm{\sigma}\rangle(\bm{k}) = \frac{\bm{B}^+_{\bm{k}}+\bm{B}^-_{\bm{k}}}{2\vert\bm{B}_{\bm{k}}\vert}.
\end{equation}
Hereafter, it is straightforward to obtain the  expectation values of the spin operators using Eq.(\ref{eq:spin_vector}) which give
\begin{subequations}
\begin{equation}
\langle\sigma_x\rangle = \frac{v_zv_{\|}k_y\cos\theta}{v_3\sqrt{v_1^2k_1^2+v_{\|}^2k_y^2}},
\label{eq:sx}
\end{equation}
\begin{equation}
\langle\sigma_y\rangle = \frac{-v_zv_{\|}k_1}{v_3\sqrt{v_1^2k_1^2+v_{\|}^2k_y^2}},
\label{eq:sy}
\end{equation}
\begin{equation}
\langle\sigma_z\rangle = 0,
\end{equation}
\label{eq:spinexpectation}
\end{subequations}
which is shown in Fig.(\ref{fig:spintexture}).

\begin{figure}[!h]
\begin{center}
\includegraphics[width=1.\columnwidth]{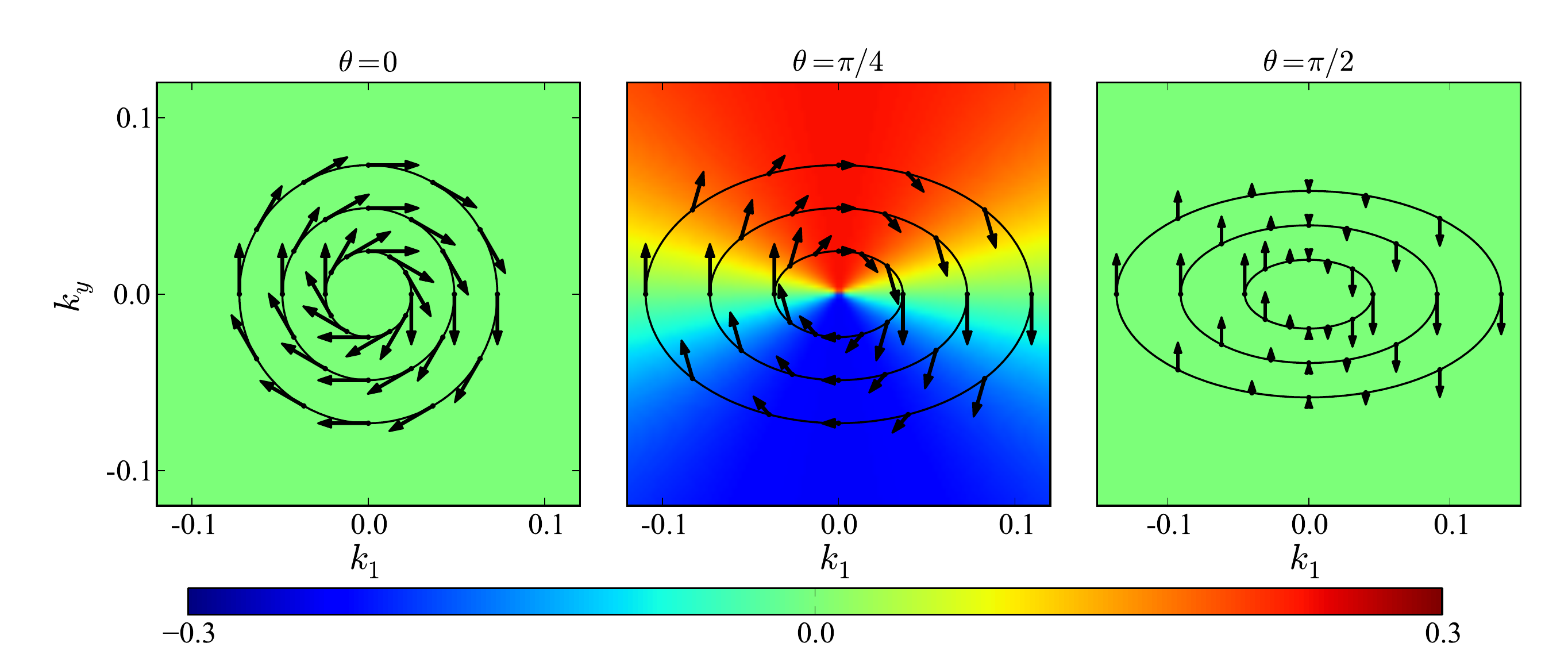}
\end{center}
\caption{The spin texture for three different surfaces states corresponding to $\theta=0, \pi/4, \text{ and } \pi/2$ are plotted in the momentum space. It should be noted that the axes of the figures are $k_1$ and $k_y$ (in units of $\text{\AA}^{-1}$) which are the local in-plane axes, and the vectors show the in-plane spin components $\sigma_1$ and $\sigma_y$. The density plot in the background shows the out-of-plane component of the magnetization $\sigma_3$. }
\label{fig:spintexturevector}
\end{figure}

Note that the expectation values of the spin operators flip their signs for valence band.
Also, note that the spin on any arbitrary surface always lies in the $x$-$y$ plane defined in the global coordinate system. Therefore, for an arbitrary surface, which is not parallel to the $x$-$y$ plane, the spin can have a component perpendicular to the surface and this is a signature of absence of a perfect spin-momentum locking in contrary to the case for the $\theta=0\text{ and }\pi$ surfaces. This becomes even more evident if the expectation value of spin is looked at in the local surface dependent coordinate system, where the surface is described by the $k_1$-$k_y$ plane and $k_3$ points perpendicular to the surface, as shown in Fig.(\ref{fig:spintexturevector}). In this local coordinate system $\langle\sigma_1\rangle_{\theta}=\langle\sigma_x\rangle\cos \theta$ and $\langle\sigma_3\rangle_{\theta}=\langle\sigma_x\rangle\sin \theta$.

Also, the magnitude of the spin $\vert\langle\bm{\sigma}\rangle\vert = \sqrt{\langle\sigma_x\rangle^2+\langle\sigma_y\rangle^2+\langle\sigma_z\rangle^2}$  has a texture in the momentum space which depends on the surface. This is an additional feature which comes along with imperfect spin-momentum locking for surfaces with $\theta \neq 0, \pi$.  Interesting features that appear in the spin texture for an arbitrary surface of the TI are shown in Fig.~(\ref{fig:spintexture}). The third column of Fig.~(\ref{fig:spintexture}) shows that for an arbitrary surfaces the magnitude of the spin on the Fermi surface is not a constant.

The imperfection in the spin-momentum locking is better highlighted if one calculates the spin-momentum locking angle ($\theta_{\text{L}}$). It turns out that it is a constant for the surface perpendicular to the crystal growth axis, whereas for arbitrary surfaces, $\theta_{\text{L}}$  has a texture in the momentum space. Using the relation $\cos\theta_{\text{L}}(\bm{k}) = \langle\bm\sigma\rangle(\bm{k})\cdot\bm{\hat{k}}/\vert \langle\bm{\sigma}\rangle(\bm{k})\vert$, and the expressions in Eqs.~(\ref{eq:sx}) and (\ref{eq:sy}), one can show that
\begin{equation}
\theta_{\text{L}}(\bm{k})=\cos^{-1}\left(\frac{-\sin^2\theta\sin\delta_{\bm{k}}\cos\delta_{\bm{k}}}{\sqrt{\cos^2\theta\sin^2\delta_{\bm{k}} + \cos^2\delta_{\bm{k}}}}\right),
\label{eq:thetal}
\end{equation}
where $\delta_{\bm{k}}$ parameterizes the momentum on the surface as $k_1 = k_p\cos\delta_{\bm{k}}$ and $k_y=k_p\sin\delta_{\bm{k}}$ with $k_p$ being the magnitude of components of the momentum in the plane.
 As can be seen from Eq.~(\ref{eq:thetal}) and Fig.~(\ref{fig:spintexturevector}) the spin always points perpendicular to the momentum for $\theta=0$, however as one begins to take arbitrary surfaces($\theta \neq 0$), the spin-momentum locking angle begins to develop a dependence on the momentum. 
 
\bibliography{references}

\end{document}